\documentclass[12pt]{article}

\usepackage{amsmath}
\usepackage{amssymb}
\usepackage{graphicx}
\usepackage{subfigure}
\usepackage{rotating}
\usepackage{color}
\usepackage{amsthm}
\usepackage{authblk}
\usepackage{natbib}
\usepackage{threeparttable}
\usepackage{pdflscape}
\usepackage{booktabs}
\usepackage{multirow}
\usepackage{accents}
\usepackage{longtable}
\usepackage{setspace}
\usepackage{url}
\usepackage[ruled]{algorithm2e}
\usepackage{algorithmic}
\usepackage{hyperref}
\usepackage{pdfpages}
\usepackage{setspace}

\begin{document}

\title{VIMCO: Variational Inference for Multiple Correlated Outcomes in Genome-wide Association Studies}
\author[1, 4]{Xingjie Shi}
\affil[1]{Department of Statistics, Nanjing University of Finance and Economics, Nanjing}
\author[2]{Yuling Jiao}
\affil[2]{School of Statistics and Mathematics, Zhongnan University of Economics and Law, Wuhan}
\author[3]{Yi Yang}
\affil[3, 4]{School of Statistics and Management, Shanghai University of Finance and Economics, Shanghai}
\author[4]{Ching-Yu Cheng}
\affil[4]{Centre for Quantitative Medicine, Duke-NUS Medical School, Singapore}
\author[5]{Can Yang}
\affil[5]{Department of Mathematics, Hong Kong University of Science and Technology, Hong Kong}
\author[4]{Xinyi Lin\footnote{To whom correspondence should be addressed, equal contribution.}}
\author[4]{Jin Liu$^\ast$}

\maketitle
\def\cal{\mathcal}
\def\ep{\varepsilon}
\def\ba{{\boldsymbol a}}
\def\bA{{\boldsymbol A}}
\def\cA{{\cal A}}
\def\hA{\widehat{A}}
\def\tba{\widetilde{\boldsymbol a}}
\def\bb{{\boldsymbol b}}
\def\hb{\hat{b}}
\def\hbb{\hat{\boldsymbol b}}
\def\tbb{\tilde{\boldsymbol b}}
\def\barb{\bar{b}}
\def\bbb{\overline{\bb}}
\def\Bbar{\overline{B}}
\def\bB{{\boldsymbol B}}
\def\brB{\boldsymbol{\mathrm B}}
\def\cB{{\cal B}}
\def\hcB{{\widehat{\cal B}}}
\def\bC{{\boldsymbol C}}
\def\bD{{\boldsymbol D}}
\def\rD{{\mathrm D}}
\def\bd{{\boldsymbol d}}
\def\bE{{\boldsymbol E}}
\def\be{{\boldsymbol e}}
\def\cE{{\cal E}}
\def\rE{{\mathrm E}}
\def\brE{\boldsymbol{\mathrm E}}
\def\bff{{\boldsymbol f}}
\def\hf{\widehat{f}}
\def\cF{{\cal F}}
\def\bg{{\boldsymbol g}}
\def\bG{{\boldsymbol G}}
\def\cG{{\cal G}}
\def\bh{{\boldsymbol h}}
\def\cH{{\cal H}}
\def\bI{{\boldsymbol I}}
\def\rI{\mathrm I}
\def\bell{{\boldsymbol \ell}}
\def\cL{{\cal L}}
\def\tM{\widetilde{M}}
\def\cM{{\cal M}}
\def\cN{{\cal N}}
\def\bO{{\boldsymbol O}}
\def\cO{{\cal O}}
\def\bp{{\boldsymbol p}}
\def\rP{{\mathrm P}}
\def\bP{{\boldsymbol P}}
\def\bbP{{\mathbb P}}
\def\tP{\widetilde{P}}
\def\bQ{{\boldsymbol Q}}
\def\bfr{{\boldsymbol r}}
\def\tr{\tilde{r}}
\def\hr{\hat{r}}
\def\hbr{\hat{\boldsymbol r}}
\def\tbr{\tilde{\boldsymbol r}}
\def\bs{{\boldsymbol s}}
\def\hs{\widehat{ s}}
\def\cS{{\cal S}}
\def\bt{{\boldsymbol t}}
\def\bT{{\boldsymbol T}}
\def\bu{{\boldsymbol u}}
\def\hbu{\widehat{\boldsymbol u}}
\def\bU{{\boldsymbol U}}
\def\rU{{\mathrm U}}
\def\tu{\tilde{u}}
\def\bv{{\boldsymbol v}}
\def\bV{{\boldsymbol V}}
\def\bw{{\boldsymbol w}}
\def\bW{{\boldsymbol W}}
\def\tw{\tilde{w}}
\def\bx{{\boldsymbol{x}}}
\def\tx{\widetilde{x}}
\def\brx{\boldsymbol{\mathrm x}}
\def\brX{\boldsymbol{\mathrm X}}
\def\tbx{\widetilde{\boldsymbol{x}}}
\def\bX{{\boldsymbol X}}
\def\cX{{\cal X}}
\def\tX{\widetilde{X}}
\def\ty{\tilde{y}}
\def\by{{\boldsymbol y}}
\def\bY{{\boldsymbol Y}}
\def\bry{\boldsymbol{\mathrm y}}
\def\brY{\boldsymbol{\mathrm Y}}
\def\tby{\tilde{\boldsymbol y}}
\def\tY{\widetilde{Y}}
\def\hy{\hat{y}}
\def\byhat{{\hat {\boldsymbol y}}}
\def\bry{\boldsymbol{\mathrm y}}
\def\tY{\widetilde{Y}}
\def\cY{{\cal Y}}
\def\Ybar{\overline{Y}}
\def\bz{{\boldsymbol z}}
\def\bZ{{\boldsymbol Z}}
\def\cZ{{\cal Z}}
\def\tz{\tilde{z}}
\def\tZ{\widetilde{Z}}
\def\brho{{\boldsymbol{\rho}}}
\def\bzero{{\boldsymbol 0}}
\def\eps{\epsilon}
\def\veps{\varepsilon}
\def\bveps{\boldsymbol{varepsilon}}
\def\tveps{\widetilde{\varepsilon}}
\def\tbveps{\widetilde{\boldsymbol{\varepsilon}}}
\def\Ghat{\widehat{G}}
\def\argmax{\mathop{\rm argmax}}
\def\argmin{\mathop{\rm argmin}}
\def\real{\mathop{{\rm I}\kern-.2em\hbox{\rm R}}\nolimits}
\def\diag{\mbox{diag}}

\def\sgn{\hbox{sgn}}
\def\tr{\hbox{tr}}
\def\Var{\hbox{Var}}
\def\vect{\hbox{vec}}
\def\Cov{\hbox{Cov}}
\def\Rem{\hbox{Rem}}

\def\whbeta{\widehat{\beta}}
\def\hbeta{\hat{\beta}}
\def\whtheta{\widehat{\theta}}
\def\htheta{\hat{\theta}}
\def\whF{\widehat{F}}
\def\hF{\hat{F}}
\def\Dnu{\Delta_{\nu}}
\def\median{\mbox{median}}
\def\sign{\mbox{sign}}
\def\trace{\mbox{trace}}
\def\Pr{\mbox{Pr}}
\def\KL{\mbox{KL}}
\def\fdr{\mbox{fdr}}
\def\bone{{\boldsymbol 1}}
\def\bzero{{\boldsymbol 0}}
\def\balpha{\boldsymbol \alpha}
\def\btheta{\boldsymbol \theta}
\def\bbeta{\boldsymbol \beta}
\def\hbbeta{\hat{\boldsymbol \beta}}
\def\hsbbetan{\hat{\boldsymbol \beta_n^*}}
\def\tbeta{\tilde{\beta}}
\def\tbbeta{\tilde{\boldsymbol \beta}}
\def\bdelta{\boldsymbol \delta}
\def\bata{\boldsymbol \eta}
\def\bgamma{\boldsymbol \gamma}
\def\lam{\lambda}
\def\blam{\boldsymbol \lambda}
\def\hmu{\widehat{\mu}}
\def\bmu{\boldsymbol \mu}
\def\bnu{\boldsymbol \nu}
\def\hphi{\widehat{\phi}}
\def\drho{\dot{\rho}}
\def\hsigma{\widehat{\sigma}}
\def\ttheta{\widetilde{\theta}}
\def\hbtheta{\widehat{\boldsymbol \theta}}
\def\bveps{\boldsymbol \varepsilon}
\def\tbveps{{\tilde\bveps}}
\def\bxi{\boldsymbol \xi}
\def\txi{\tilde{\xi}}
\def\tzeta{\tilde{\zeta}}

\begin{abstract}
{
	In Genome-Wide Association Studies (GWAS) where multiple correlated traits have been measured on participants, a joint analysis strategy, whereby the traits are analyzed jointly, can improve statistical power over a single-trait analysis strategy. 
	There are two questions of interest to be addressed when conducting a joint GWAS analysis with multiple traits. The first question examines whether a genetic loci is significantly associated with any of the traits being tested. The second question focuses on identifying the specific trait(s) that is associated with the genetic loci. Since existing methods primarily focus on the first question, this paper seeks to provide a complementary method that addresses the second question. We propose a novel method, Variational Inference for Multiple Correlated Outcomes  (VIMCO), that focuses on identifying the specific trait that is associated with the genetic loci, when performing a joint GWAS analysis of multiple traits, while  accounting for correlation among the multiple traits.
	We performed extensive numerical studies and also applied VIMCO to analyze two datasets. The numerical studies and real data analysis demonstrate that VIMCO improves statistical power over single-trait analysis strategies when the multiple traits are correlated and has comparable performance when the  traits are not correlated.
	
}
\end{abstract}

\section{Introduction}
Genome-wide association studies (GWAS) conducted in the last decade have provided valuable insights into the genetic architecture underlying complex traits. GWAS are generally performed by analyzing individual traits, although multiple related traits are often collected, and in some cases these traits may reflect a common condition  \citep{kim2009multivariate}. 
In a GWAS where multiple traits have been measured on the study individuals, a joint analysis strategy whereby the multiple traits are analyzed jointly, offers improved statistical power when compared to a single-trait analysis strategy, when a genetic variant is associated with one or more correlated phenotypes  \citep{korte2012mixed, solovieff2013pleiotropy}. A joint analysis strategy may also be useful when the different phenotypes characterize the same underlying trait.

There are typically two questions of interest to be addressed in a joint GWAS analysis of multiple traits. The first question addresses whether a genetic loci is significantly associated with any of the (correlated) phenotypes being tested. The second question, focuses on the identification of the phenotype(s) that is/are associated with the genetic loci. When multiple continuous traits and genotype data are available from the same study individuals, a joint GWAS analysis that addresses the first question can be performed using linear mixed models based methods \citep{casale2016multivariate}. Examples of  linear mixed models based methods include the multi-trait mixed models (MTMM) proposed by \cite{korte2012mixed} and the multivariate linear mixed models (mvLMM) implemented in the Genome-wide Efficient Mixed Model Association (GEMMA) software \citep{zhou2014efficient}. A central limitation of these methods is that they cannot address the second question which seeks to identify the tested trait(s) that is/are associated with the genetic loci.  Multi-trait variable selection methods based on penalization have also been proposed in this context \citep{rothman2010sparse, liu2016analyzing}. However, they cannot be applied to assess associations while controlling the error rate \citep{carbonetto2012scalable}.

In this paper, we propose a novel method, Variational Inference for Multiple Correlated Outcomes  (VIMCO), for joint analysis of multiple traits in GWAS that addresses the second  question. VIMCO is applicable when individual-level data on multiple traits and genotype are available on the same study individuals. Our proposed method, can be viewed as a complementary method to the widely-used mvLMM implemented in the GEMMA package, in that it addresses a different but related scientific question and allows one to identify the specific trait that is associated with the genetic loci when performing a joint analysis of multiple traits. A variational Bayesian expectation-maximization (VBEM) algorithm is used to ensure computational efficiency. Through extensive numerical studies and real data analyses, we demonstrate that our proposed approach offers improved statistical power when compared to existing single-trait analysis strategies. The remainder of this article is organized as follows. In Section 2, we describe the model and algorithm that VIMCO uses to perform  joint analysis of multiple traits. We then illustrate the performance of VIMCO using numerical simulations and real data analyses in Section 3 and conclude with a discussion in Section 4.

\section{VIMCO}
\subsection{Model}
In this section, we describe the notation and model  used for joint modeling of multiple traits in a GWAS using VIMCO.
Consider $K$ continuous phenotypes/traits, $Y_1, \dots, Y_K$, that are measured on $N$ individuals, where $Y_k$ is a $N\times 1$ vector for $k=1,\cdots,K$.
Assume that the genome-wide genotype data consists of $p$ SNPs given by, $X_1, \dots, X_p$ where $X_j$ is a $N\times 1$ vector for $j=1,\cdots,p$.
Denote $\brX=[X_1, \dots, X_p]\in\mathbb R^{N\times p}$ and $\brY = [Y_1, \dots, Y_K]\in\mathbb R^{N\times K}$. 
Without loss of generality, we assume that both the phenotypes and genotypes have been centered.
We consider the following multivariate linear model:
\begin{equation}\label{lm}
	\brY = \brX\brB + \brE,
\end{equation}
where $\brB=[\bbeta_1,\dots,\bbeta_p]^\top\in \mathbb R^{p\times K}$, $\bbeta_j=(\beta_{j 1},\dots,\beta_{jK})^\top$, $j=1,\dots,p$, and $\brE=[e_1^\top,\dots,e_N^\top]^\top\in \mathbb R^{N\times K}$. We assume that $e_n \sim \cN(\bzero, \Theta^{-1})$, where $\Theta$ is the precision matrix of $\brE$ with dimensionality $K\times K$ for individuals $n=1, \cdots, N$. The entries in $\Theta$ are denoted by $\theta_{st}, s, t=1,\dots, K$. Under this model, the correlation in the traits (conditional on the genotypes) is modeled by the off-diagnoal terms in $\Theta^{-1}$.

We are interested in identifying  genetic variants that are associated with one or more traits, which corresponds to the identification of nonzero entries in the matrix $\brB$. 
We consider a spike-slab prior for $\brB$ and parameterize $\brB$ as a product of latent variables. Specifically, we assume that $\bbeta_{j}=\bgamma_{j}\circ\tbbeta_{j}=(\gamma_{j1}\tilde\beta_{j1}, \dots, \gamma_{jK}\tilde\beta_{jK})^\top$, where  $\circ$ denotes the element-wise product and  
\[
\tbeta_{jk}\sim\cN(0,\sigma_{\beta_k}^2),  ~\gamma_{jk}\sim a_k^{\gamma_{jk}}(1-a_k)^{1-\gamma_{jk}}.
\]
Let $\Phi=\{a_1, \dots, a_K, \sigma_{\beta_1}^2,\dots,\sigma_{\beta_K}^2,\Theta\}$ be the collection of (unknown) model parameters. Accordingly, our probabilistic model can be reparameterized as:
\begin{align*}
	& \Pr(\brY,\tbbeta,\bgamma|\brX;\Phi) 
	=   ~ \Pr(\brY|\brX,\tbbeta,\bgamma;\Phi) \Pr(\tbbeta,\bgamma|\brX;\Phi) \\
	=  & \prod_{n=1}^N \cN(\sum_{j=1}^pX_{nj} \left(\bgamma_j \circ \tbbeta_{j}\right), \Theta^{-1}) \prod_{j=1}^p\prod_{k=1}^K\left[a_k^{\gamma_{jk}}(1-a_k)^{1-\gamma_{jk}}  \cN(0,\sigma_{\beta_k}^2) \right].
\end{align*}

With this reparameterization, to identify genetic variants that are associated with one or more traits, we need to compute the posterior distribution of the latent variables $(\tbbeta,\bgamma)$:
\begin{align}
	\Pr(\tbbeta,\bgamma|\brY, \brX;\Phi) &= \frac{\Pr(\brY,\tbbeta,\bgamma|\brX;\Phi) }{\Pr(\brY|\brX;\Phi)}\nonumber\\&=\frac{\Pr(\brY,\tbbeta,\bgamma|\brX;\Phi) }{\sum_{\bgamma}\int_{\tbbeta} \Pr(\brY,\tbbeta,\bgamma|\brX;\Phi) d\tbbeta}.\label{eqn:P}
\end{align}



\subsection{VBEM algorithm}
In this section, we describe the algorithm used for computation of the model parameters and the posterior distribution of the latent variables $(\tbbeta,\bgamma)$ in VIMCO. Exact computation of the posterior distribution \eqref{eqn:P} is computationally intensive due to the denominator which requires marginalizing over the latent variables $(\tbbeta,\bgamma)$.  
To overcome this computational intractability,  we derive a computationally efficient VBEM algorithm \citep{Bishop2016pattern} to obtain an approximation for the posterior distribution \eqref{eqn:P}.
The key idea of the VBEM algorithm is to approximate our computationally intractable posterior distribution \eqref{eqn:P}, with an approximating distribution, which is computationally tractable. 
The VBEM algorithm proceeds by specifying a family of (variational) distributions, which are parameterized by variational parameters, and choosing the optimal approximating distribution from this family by minimizing the KL divergence between the approximating distributions and our true posterior distribution \eqref{eqn:P}  \citep{blei2017variational}. The KL divergence can be viewed as a measure of how different the approximating distribution is from our true posterior distribution \eqref{eqn:P}. The optimal variational distribution is then used as an approximation for the true posterior distribution \eqref{eqn:P} \citep{Bishop2016pattern}. 

Let $q(\tbbeta,\bgamma)$ be a candidate approximating distribution of our true posterior distribution \eqref{eqn:P}, and let $\rE_q$ denote the  expectation taken with respect to $q(\tbbeta,\bgamma)$. We can decompose the logarithm of the marginal likelihood as
\begin{equation}\label{eqn:logML}
	\begin{split}
		\log p(\brY|\brX;\Phi) 
		& = \cL_q 
		+ \KL\left(q || \Pr(\tbbeta,\bgamma|\brY, \brX;\Phi) \right),
	\end{split}
\end{equation}
where 
\begin{equation*}
	\begin{split}
		\cL_q &= \rE_q \log\left[\frac{ \Pr(\brY,\tbbeta,\bgamma|\brX;\Phi) }{q(\tbbeta,\bgamma)}\right]\\
		\KL\left(q || \Pr(\tbbeta,\bgamma|\brY, \brX;\Phi) \right)  &= \rE_q\log \left[ \frac{q(\tbbeta,\bgamma)}{ \Pr(\tbbeta,\bgamma|\brY, \brX;\Phi) }\right].
	\end{split}
\end{equation*}
Minimizing the KL divergence with respect to the approximating distribution $q(\tbbeta,\bgamma)$ is equivalent to maximizing the evidence lower bound (ELBO) $\cL_q$. If we allow any possible choice for $q(\tbbeta,\bgamma)$, then the Kullback-Leibler divergence  $\KL\left(q|| \Pr(\tbbeta,\bgamma|\brY, \brX;\Phi) \right)$ is zero if and only if $q(\tbbeta,\bgamma)$ is identical to our true posterior distribution \eqref{eqn:P} almost surely.

In order for the algorithm to be applicable to GWAS data, we require a family of distributions that is sufficiently flexible to accurately approximate the true posterior distribution, while being computationally tractable. We propose using a mean field variational family  \citep{logsdon2010variational}, which contains distributions $q(\tbbeta,\bgamma)$ of the form
\[
q(\tbbeta,\bgamma) = \prod_{j}  \prod_{k} \left[q(\tbeta_{jk}, \gamma_{jk}) \right].
\]
The factorization used in the approximating mean field variational family of distributions makes the VBEM algorithm computationally efficient. The approximation is expected to perform best when the SNPs are independent and in the absence of pleiotropy. Our numerical studies demonstrate that this approximation is sufficient in the presence of moderate SNP correlation and moderate levels of pleiotropy. When the SNPs are highly correlated, SNP pruning can be used to retain a subset of SNPs that are moderately correlated, as was done in the real data analysis.

Given this variational family of distributions, the optimal variational distribution $q^\ast(\tbeta_{jk},\gamma_{jk})$ that maximizes the ELBO $\cL_q$ has the form  \citep{Bishop2016pattern}
\begin{equation}\label{eqn:qm}
	\log q^\ast(\tbeta_{jk},\gamma_{jk}) = \rE_{(j', k')\neq (j, k)}\left[\log\Pr(\brY,\tbbeta,\bgamma|\brX;\Phi)\right] + \text{constant},
\end{equation}
where the expectation is taken with respect to all  other factors $q(\tbeta_{j'k'},\gamma_{j'k'})$ for $(j', k')\neq (j, k)$. After some derivations (details are provided in  Supplementary A), the optimal variational posterior distribution is given by:
\begin{equation}\label{eqn:vd}
	\prod_{j}  \prod_k\left[ \alpha_{jk}^{\gamma_{jk}}(1-\alpha_{jk})^{1-\gamma_{jk}}\cN(\mu_{jk},s_{jk}^2)^{\gamma_{jk}} \cN(0,\sigma_{\beta_k}^2)^{1-\gamma_{jk}}\right], \nonumber
\end{equation}
where the   variational parameters ($\mu_{jk}$, $s_{jk}^2$, $\alpha_{jk}$) are given by:
\begin{equation}
	\begin{split}
		\mu_{jk}  &=\frac{\sum_t\theta_{kt} X_j^\top [Y_t - \sum_{j'\neq j}\alpha_{j't}\mu_{j't}X_{j'}] - \sum_{t\neq k}\theta_{kt}\alpha_{jt}\mu_{jt} \|X_j\|^2}{\theta_{kk}\|X_j\|^2+\frac{1}{\sigma_{\beta_k}^2}},\\
		s_{jk}^2 &= \frac{1}{\theta_{kk}\|X_j\|^2+\frac{1}{\sigma_{\beta_k}^2}},\\
		\alpha_{jk} & \equiv q(\gamma_{jk}=1) = \frac{1}{1+\exp\left(-\log\frac{a_k}{1-a_k}  + \frac{1}{2} \left(\frac{\mu_{jk}^2}{s_{jk}^2}+\log\frac{s_{jk}^2}{\sigma_{\beta_k}^2}\right)\right)}. \label{eqn:EStep1}
	\end{split}
\end{equation}

To solve for the variational parameters ($\mu_{jk}$, $s_{jk}^2$, and $\alpha_{jk}$) and model parameters ($\Phi$), the VBEM algorithm iterates between  two optimization (expectation and maximization) steps until convergence. In the expectation step, we optimize the ELBO $\cL_q$ with respect to the variational parameters ($\mu_{jk}$,  $s_{jk}^2$ and  $\alpha_{jk}$), while holding the model parameters fixed, i.e. compute variational parameters using Equation \eqref{eqn:EStep1}. In the maximization step,  we optimize the ELBO $\cL_q$ with respect to the model parameters $\Phi$  while holding the variational parameters fixed. With the optimal variational distribution, the ELBO  $\cL_q$
can be evaluated in a closed form (details  in  Supplementary A): 
\begin{equation} 
	\begin{split}
		\cL_q  = 
		& - \frac{1}{2}\sum_s\sum_t\theta_{st}(Y_s - \sum_jX_{j}\alpha_{js}\mu_{js})^\top(Y_t - \sum_jX_{j}\alpha_{jt}\mu_{jt}) \\
		& -\frac{1}{2}\sum_s\theta_{ss}\sum_jX_j^\top X_j[\alpha_{js}(\mu_{js}^2+s_{js}^2) - \alpha_{js}^2\mu_{js}^2]\\
		& - \sum_j\sum_k\left[\alpha_{jk}\log\frac{\alpha_{jk}}{a_k} + (1-\alpha_{jk})\log\frac{1-\alpha_{jk}}{1- a_k}\right]+  \frac{N}{2}\log|\Theta| \\
		& 
		+ \frac{1}{2}\sum_{j}\sum_{k} \alpha_{jk} \left(1 + \log\frac{s_{jk}^2}{\sigma_{\beta_k}^2} -\frac{\mu_{jk}^2+s_{jk}^2}{\sigma_{\beta_k}^2} \right) + \text{const.}
	\end{split}
\end{equation}

By taking partial derivatives of the ELBO $\cL_q$ with respect to the model parameters and setting them to zero, we can solve for the model parameters and obtain the update equations for the maximization step:
\begin{equation}\label{eqn:MStep}
	\begin{split}
		a_k = & \frac{\sum_j \alpha_{jk}}{p},\\
		\sigma_{\beta_k}^2  =&   \frac{\sum_j \alpha_{jk}(\mu_{jk}^2+s_{jk}^2)}{\sum_j \alpha_{jk}},\\
		(\Theta^{-1})_{kk} =  & \frac{ \|Y_{k}-\sum_jX_{j}\alpha_{jk}\mu_{jk}\|^2}{N} \\
		& +  \frac{ \sum_j \| X_j\|^2 \left[\alpha_{jk}(\mu_{jk}^2+s_{jk}^2)-\alpha_{jk}^2\mu_{jk}^2\right]}{N}, \\
		(\Theta^{-1})_{kt} =& \frac{(Y_{k}-\sum_jX_{j} \alpha_{jk}\mu_{jk})^\top(Y_{t}-\sum_jX_{j} \alpha_{jt}\mu_{jt}) }{N}.
	\end{split}
\end{equation}
The VBEM algorithm iterates between the expectation  (Equation \eqref{eqn:EStep1}) and maximization (Equation \eqref{eqn:MStep}) steps until convergence. 
Further details on the VBEM algorithm are provided in Supplementary A.

\subsection{Inference}
With the estimated variational parameters ($\mu_{jk}$, $s_{jk}^2$ and $\alpha_{jk}$) and model parameters ($\Phi$), the posterior probability $\Pr(\gamma_{jk}=1|\brY,\brX; \Phi)$ of whether genetic variant $j$ is associated with trait $k$ can be estimated by $\widehat{\alpha}_{jk}$, and the  local false discovery rate (lfdr) can be estimated by $1 - \widehat{\alpha}_{jk}$. Statistical inference can be conducted by identifying SNP-trait associations while controlling the global FDR at a fixed value.
Specifically, given a cutoff for the global FDR, the cutoff $\xi$ for the lfdr can be computed from $\text{global FDR} = \frac{\sum_j\sum_k \text{lfdr}_{jk} \mathbb I(\text{lfdr}_{jk} \leq \xi)}{\sum_j\sum_k \mathbb I(\text{lfdr}_{jk} \leq \xi)}$ \citep{newton2004detecting}.

\section{Results}
\subsection{Simulations}
We conducted numerical simulations to evaluate the performance of VIMCO. We considered the scenario where we have $K=4$ continuous traits and $p=10,000$ SNPs that were measured on $N=5,000$ individuals.
For each individual, the $p$ genotypes were simulated by first generating a $p\times 1$ multivariate normal distribution assuming auto-regressive (AR) correlation with parameter $\rho_x$. We then discretized each variable to a trinary variable  $(0, 1, 2)$ by assuming  Hardy-Weinberg equilibrium and with a minor allele frequency randomly selected from  a $\text{uniform}[0.05, 0.5]$ distribution. The genotype correlation was varied at $\rho_x= 0.2, 0.5, 0.8$. 
To generate the coefficient matrix $\brB$, for each trait, we randomly selected 1\% of the SNPs to be associated with the trait, where the effect sizes were generated from a standard normal distribution. To allow for pleiotropy where a SNP can be associated with more than one trait, we varied the proportion of causal SNPs that was associated with more than one trait. Let $g = \frac{\sum_j \mathbb I (\sum_k \gamma_{jk} \geq 2) }{\sum_j\sum_k \gamma_{jk}}$. The expectation of $g$ was varied at $0, 0.15, 0.3$, where increasing $g$ reflects increasing pleiotropy. The error matrix $\brE$ was generated with rows drawn independently from a multivariate normal distribution, with auto-regressive (AR) correlation parameter $\rho_e= 0.2, 0.5, 0.8$. A larger value of $\rho_e$ implies a higher correlation between the traits. Each error variance was adjusted according to the prespecified heritability of $h^2=0.3$. 

\begin{figure*}[t]
	\centering
	\includegraphics[width=\linewidth]{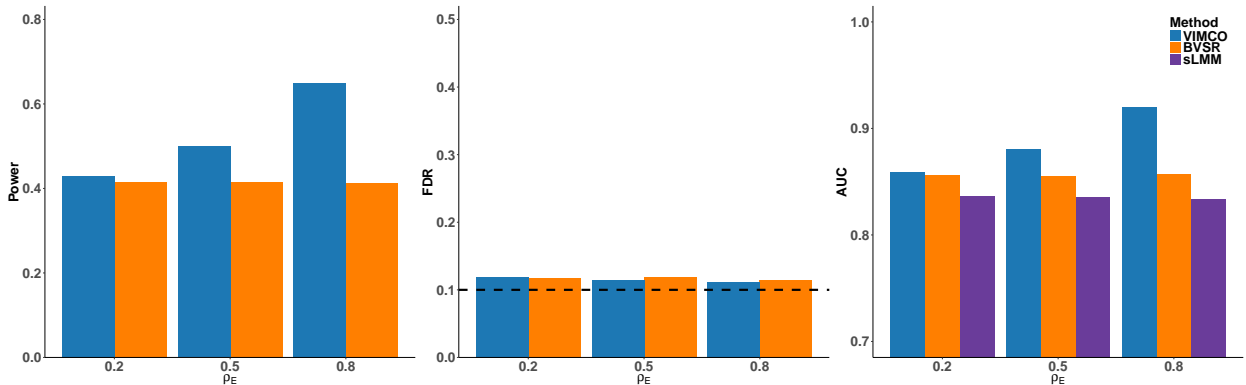}
	\caption{Simulation results for evaluating null hypothesis $H_{0a}$, for different $\rho_e$ (increasing levels of $\rho_e$ imply increasing correlation between the traits) and genotype correlation parameter $\rho_x=0.8$. Left panel: power of VIMCO and BVSR; middle panel: empirical global false discovery rate (FDR) of VIMCO and BVSR at a nominal 0.1 level; right panel: AUC of VIMCO, BVSR, and sLMM.
	}
	\label{fig:Harhox08}
\end{figure*}

To benchmark the performance of VIMCO, we considered  both bayesian (variational inference based bayesian variable selection regression (BVSR) \citep{carbonetto2012scalable}) and frequentist  (single-trait linear mixed model (sLMM) \citep{zhou2012genome}) single-trait analysis approaches. Similar to VIMCO, BVSR utilizes a VBEM algorithm, but can only be applied to single traits. sLMM was implemented using the GEMMA software package.  We compared the power of VIMCO and BVSR, by considering trait-SNP associations from all traits and SNPs, and with a global FDR controlled at $0.1$. 
The power for VIMCO and BVSR,  for the scenario where the genotypes were strongly correlated ($\rho_x=0.8$) and pleiotropy $g=0$, is shown in the left panel of Figure  \ref{fig:Harhox08}. 
When the traits showed moderate ($\rho_e=0.5$) or strong correlation ($\rho_e=0.8$), VIMCO had higher statistical power than BVSR. When the traits were weakly correlated ($\rho_e=0.2$), VIMCO had similar power as BVSR.  We also evaluated the global FDR control for both VIMCO and BVSR. Similar to earlier papers \citep{brzyski2017controlling}, when evaluating the control of global FDR, the SNPs were evaluated as a cluster of SNPs. i.e. rejections within the same linkage disequilibrium (LD) block were grouped and counted as a single rejection. As shown in the middle panel in Figure \ref{fig:Harhox08}, both VIMCO and BVSR.
had empirical false discovery rates that were close to the nominal 0.1 level. We also compared the area under the curve (AUC) of VIMCO, BVSR and sLMM (right panel of Figure  \ref{fig:Harhox08}). The AUC was evaluated  by considering trait-SNP associations for all traits and SNPs.  The AUC of VIMCO was higher than that of the single-trait approaches (BVSR and sLMM) when the traits showed moderate ($\rho_e=0.5$) or strong correlation ($\rho_e=0.8$). Simulations with lower genotype correlation $\rho_x= 0.2, 0.5$ and different levels of pleiotropy $g$ are shown in Supplementary Figures B1-B3, and give similar conclusions.

\begin{figure*}[t]
	\centering
	\includegraphics[width=\linewidth]{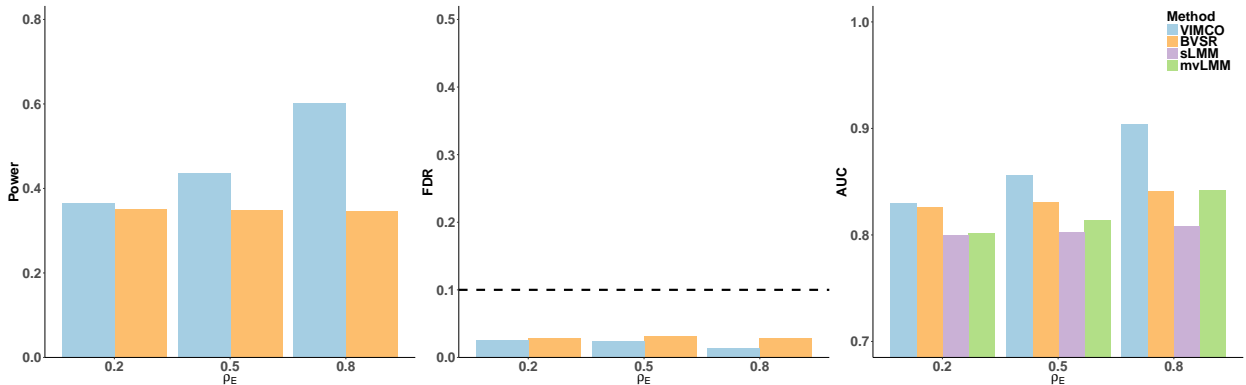}
	\caption{Simulation results for evaluating null hypothesis $H_{0b}$, for different $\rho_e$ (increasing levels of $\rho_e$ imply increasing correlation between the traits) and genotype correlation parameter $\rho_x=0.8$. Left panel: AUC of VIMCO and BVSR; middle panel: empirical global false discovery rate (FDR) of VIMCO and BVSR, at a nominal 0.1 level; right panel: AUC of VIMCO, BVSR, sLMM and mvLMM.}
	\label{fig:Hbrhox08}
\end{figure*}

As noted earlier, mvLMM  assesses a  different but related null hypothesis of whether any of the traits are associated with the genetic variants. Specifically, for the $j^{\text{th}}$ SNP, mvLMM evaluates the null hypothesis $H_{0b}: \beta_{j1}=\cdots=\beta_{jK}=0$, while VIMCO evaluates the null hypothesis $H_{0a}: \beta_{jk}=0$ for $k=1,\cdots,K$ traits separately. We evaluated the performance of VIMCO for assessing the null hypothesis $H_{0b}$. For evaluating the performance in assessing the null hypothesis $H_{0b}$, we performed an adhoc modification of VIMCO, BVSR and sLMM. 
In this ad-hoc adaptation of VIMCO, BVSR and sLMM, we rejected the null hypothesis $H_{0b}$ if $H_{0a}$ is rejected for any of the traits. We examined the power of VIMCO and BVSR while controlling the global FDR at 0.1. We also compared the  AUC of VIMCO, BVSR, sLMM and mvLMM. Results for the scenario where the genotypes were strongly correlated ($\rho_x=0.8$) and the level of pleiotropy $g=0$ are shown in Figure  \ref{fig:Hbrhox08}. Similar to results in evaluating $H_{0a}$, VIMCO improves statistical power and  has higher AUC when the traits showed moderate or strong correlation.
Interestingly for this ad-hoc adaptation of VIMCO and BVSR, the empirical FDR were conservative for the settings we considered (middle panel of Figure  \ref{fig:Hbrhox08}). We note that evaluating the null hypothesis $H_{0b}$ is not an intended use of VIMCO, and this ad-hoc adaptation of VIMCO was performed in order to provide a comparison with mvLMM. Simulations with different genotype correlation and levels of pleiotropy gave similar conclusions (Supplementary Figures B4-B6). Simulations where genotypes were sampled from real data, and SNP pruning was applied before applying the methods, gave similar conclusions (Supplementary Figures B7-B8).

\subsection{Real Data Analysis}
To illustrate the performance of our proposed method VIMCO, we analyzed two datasets. The first dataset is a GWAS of four moderately correlated lipid traits from the Northern Finland Birth Cohort 1966 (NFBC1966). The second dataset is a GWAS of three weakly correlated eye measurements from the Singapore Indian Eye (SINDI) study \citep{cheng2013nine}.

\subsubsection{NFBC1966}
The NFBC1966 dataset consists of 10 metabolic traits and 364,590 SNPs from 5,402 individuals \citep{sabatti2009genome}. 
The 10 metabolic traits include fasting lipid levels (total cholesterol (TC), high-density lipoprotein (HDL), low-density lipoprotein (LDL) and triglycerides (TG)),  inflammatory marker C-reactive protein (CRP), markers of glucose homeostasis (glucose and insulin), body mass index  and blood pressure measurements (systolic and diastolic blood pressure). Quality control of the data was performed using PLINK \citep{purcell2007plink} and GCTA \citep{yang2011gcta}. Individuals with missing-ness in any of the traits and with genotype missing call-rates $>5\%$ were excluded. 
We excluded SNPs with minor allele frequency $<1\%$, missing call-rates $>1\%$, or failed Hardy-Weinberg equilibrium. After quality control filtering and SNP pruning,  172,412 SNPs from 5,123 individuals were available for analysis. We quantile-transformed  each trait to a standard normal distribution, obtained the residuals after regressing out
the effects of sex, oral contraceptives and pregnancy status, and quantile-transformed the residuals to a standard normal distribution. 

We performed analysis on the  four lipid traits (TC, LDL, HDL and TG). The  pairwise pearson correlation for the four lipid traits are given in Supplementary Figure C1. Among the four traits, TC and LDL showed the strongest correlation (corr $=0.88$). Modest correlation was also observed among the following pairs of traits: TG and TC (corr $=0.41$), TG and LDL (corr $=0.33$), TG and HDL (corr $=-0.40$).  We applied VIMCO to perform joint analysis of the 4 traits. 
For comparison with the results from VIMCO, we also performed single-trait analyses whereby each of the  4 traits were analyzed separately, using both bayesian (variational inference based bayesian variable selection regression (BVSR) \citep{carbonetto2012scalable}) and frequentist approaches (single-trait linear mixed model (sLMM) \citep{zhou2012genome}). 

For VIMCO and BVSR, we report significant SNP-trait associations at a global FDR of $0.1$. The global FDR for VIMCO and BVSR controls for multiple testing across the multiple traits and SNPs. For sLMM, to control for multiple testing across both traits and SNPs, we report p-values for SNP-trait associations with $p$-value $<1.25\times 10^{-8}$ (we applied a Bonferroni adjustment for the 4 traits to the commonly used genome-wide significance threshold $5\times 10^{-8}$).

\begin{figure*} [t]
	\centering
	\includegraphics[width=0.9\linewidth]{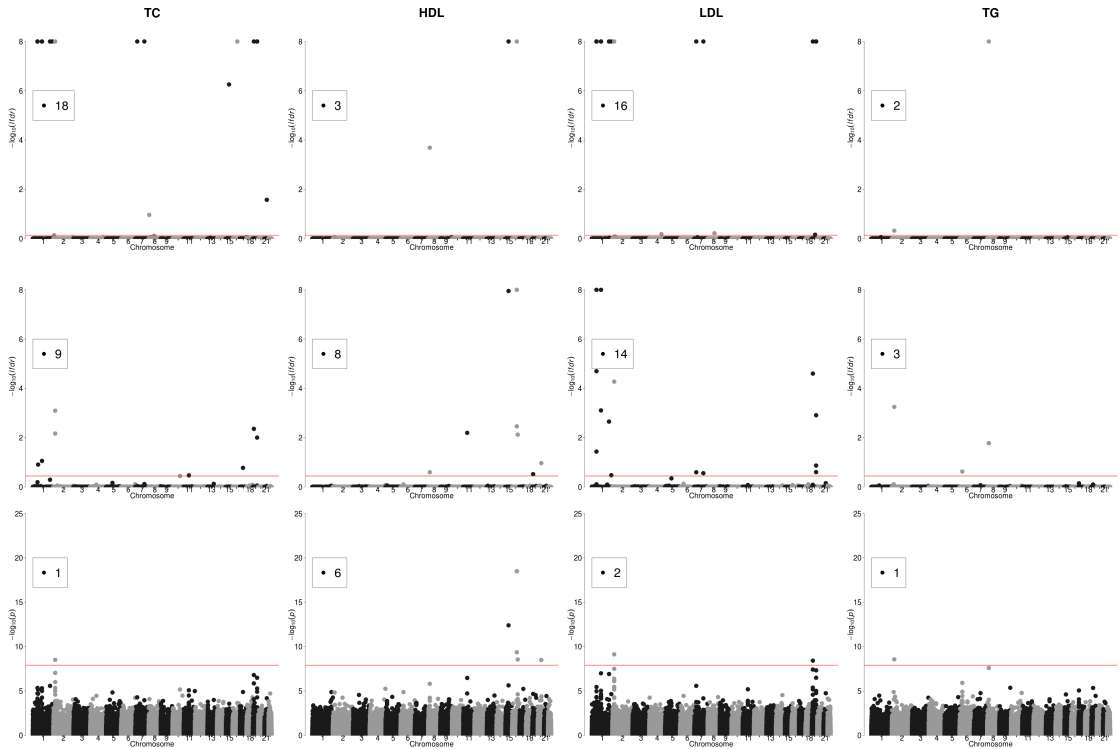}
	\caption{Manhattan plots for the analysis results of VIMCO (first row), BVSR (second row), and sLMM (last row), in  the NFBC1966 study. In the first and second row, the horizontal red lines correspond to a global FDR of $0.1$. In the last row, the horizontal red line corresponds to a $p$-value cut-off of $1.25\times 10^{-8}$. The number in the box indicates the number of SNP associations identified for each trait.}
	\label{fig:ManhNFBC}
\end{figure*}

Genomic locations of SNPs identified by VIMCO, BVSR and sLMM  are shown in Figure \ref{fig:ManhNFBC}. To control the global FDR at $0.1$, a SNP has to have a lfdr $<0.73$ and $<0.36$ for VIMCO and BVSR respectively (indicated by the  horizontal red lines in the plots). VIMCO identified a total of 39 SNP-trait associations while the single-trait analysis strategies BVSR and sLMM identified 34 and 10 SNP-trait associations, respectively. In terms of the total number of unique SNPs identified, VIMCO identified a smaller number of SNPs than BVSR; VIMCO identified 23 SNPs to be associated with at least one trait, while BVSR and sLMM identified 30 and 9 SNPs respectively. However, if we consider each trait separately, VIMCO identified more SNPs than BVSR and sLMM for traits which showed strong correlation with each other. For example, the strongest trait-trait correlation was observed between TC and LDL and for TC, VIMCO identified 18 SNPs, while BVSR and sLMM identified 9 and 1 SNPs respectively; for LDL, VIMCO identified 16 SNPs, while BVSR and sLMM identified 14 and 2 SNPs respectively.

The lfdrs and $p$-values of identified SNPs by VIMCO and BVSR are given in Supplementary Table C1. For SNPs that were identified by either VIMCO, BVSR or sLMM, we also report their p-values from a multivariate linear mixed model (mvLMM) fitted using the GEMMA package \citep{zhou2014efficient}. As noted earlier, mvLMM  assesses a different but related hypothesis of whether any of the 4 traits are associated with the genetic variants. 

Analysis of the NFBC1966 dataset was conducted on a machine with 3.0 GHz Intel Xeon CPU and 32G memory. With the BVSR estimates as initial values (1.2 hours), it took VIMCO an additional 2.5 hours to complete the full variational inference procedure.

\subsubsection{SINDI}
\begin{figure*} 
	\centering
	\includegraphics[width=0.9\linewidth]{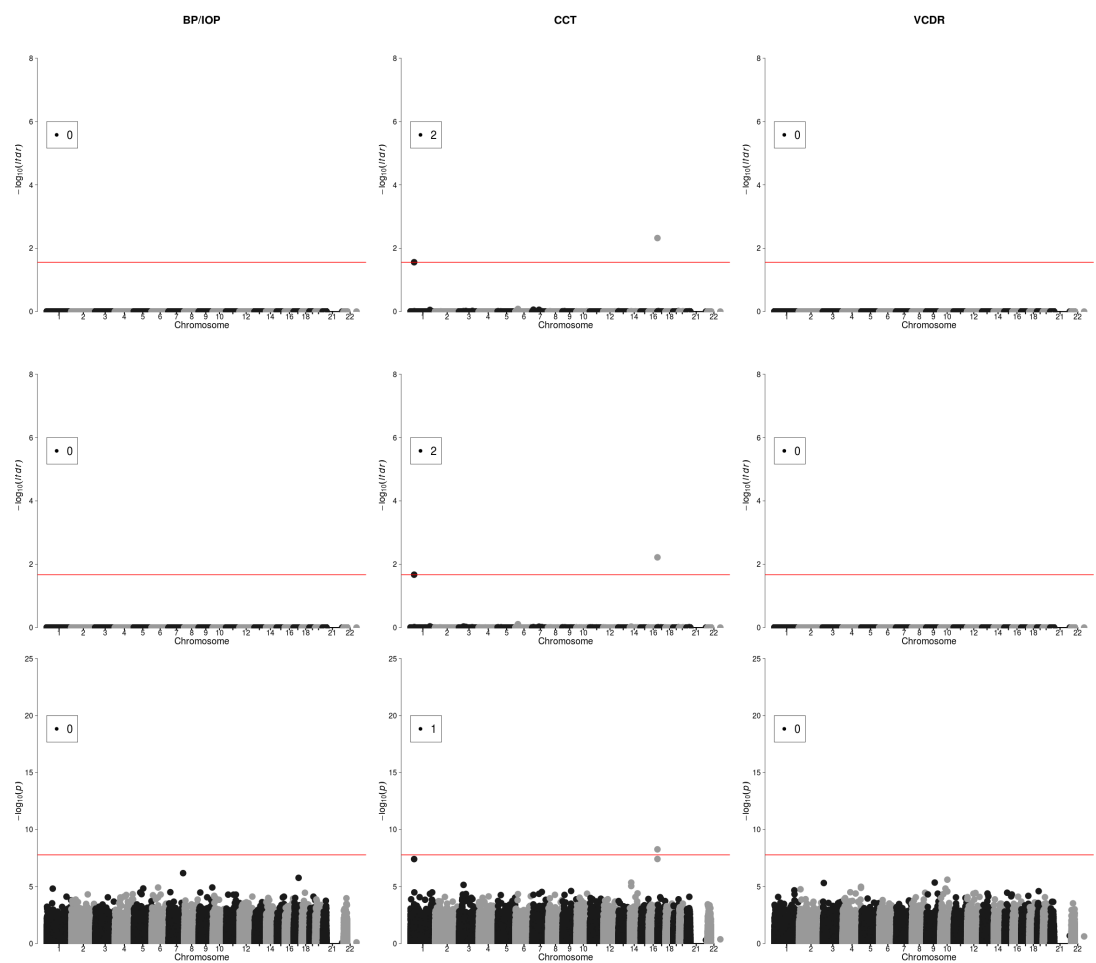}
	\caption{Manhattan plots for the analysis results of VIMCO (first row), BVSR (second row), and sLMM (last row), in the SINDI study. In the first and second row, the horizontal red lines correspond to a global FDR of $0.1$. In the last row, the horizontal red line corresponds to a $p$-value cut-off of $1.67\times 10^{-8}$. The number in the box indicates the number of SNP associations identified for each trait.}
	\label{fig:ManhSINDI}
\end{figure*}

The SINDI dataset contains three eye measurements: the ratio of blood pressure to intraocular pressure (BP/IOP), central corneal thickness (CCT) and and vertical cup-to-disc ratio (VCDR). These three traits are risk factors for glaucoma \citep{lavanya2009methodology}.
The pairwise correlation for these three traits (Supplementary Figure C2) were much weaker than those observed for the lipid traits in the NFBC1966 dataset, with the strongest correlation of 0.16 observed between BP/IOP and CCT. After quality control following previous studies \citep{cheng2013nine} and SNP pruning,  2,219 individuals and 257,736 SNPs were available for analysis. 
Locations of SNPs identified by VIMCO, BVSR and sLMM in the genome are shown in Figure \ref{fig:ManhSINDI}. In this dataset where the traits were weakly correlated, the performance for VIMCO was similar as the single trait approaches. VIMCO and BVSR identified the same 2 SNPs to be associated with CCT. Among the two associations, rs12447690 was also identified by sLMM (the $p$-value is 5.5$\times10^{-9}$). This SNP was located in the Zinc-Finger protein (ZNF469) gene, and was previously reported to be associated with CCT \citep{gao2016genome}. The lfdrs and $p$-values of identified SNPs are given in  Supplementary Table C2.

Analysis of the SINDI dataset used BVSR estimates as initial values (6.8 hours), and used an additional  1.2 hours to complete the full variational inference procedure. 

\section{Discussion}
In this article, we have proposed a novel method VIMCO that allows an investigator to identify the specific trait that is associated with the genetic loci when performing a joint GWAS analysis of multiple traits. Results from simulations and real data analyses demonstrate that  VIMCO improved statistical power when the traits were correlated and had comparable performance when the traits were not correlated, when compared with single-trait analysis strategies. Furthermore, VIMCO utilizes a computationally efficient VBEM algorithm which allows it to handle genome-wide genotype data efficiently. It is, however, not without limitations. A limitation of VIMCO is that it is not applicable when the number of traits analyzed exceeds the number of samples. With increasing interest in performing phenome-wide association studies whereby the number of phenotypes can be larger than the sample size, extending VIMCO to handle larger number of phenotypes is an avenue for future work. Additionally, VIMCO requires individual-level trait and genotype data to be collected from the same individuals. The development of a method where summary statistics from different individuals can be used for analysis instead of individual-level data is another avenue for future research. 


\section*{Funding}
This work was supported in part by grant No. 71501089 and No. 11501579 from National Natural Science Foundation of China, grants No. 22302815, No. 12316116 and No. 12301417 from the Hong Kong Research Grant Council, and grant R-913-200-098-263 and R-913-200-127-263 from the Duke-NUS Graduate Medical School, and AcRF Tier 2 (MOE2016-T2-2-029) from the Ministry of Education, Singapore.

\bibliographystyle{plain}
\bibliography{reference}


%

%


\end{document}